\documentclass{webofc}
\usepackage[varg]{txfonts}   % Web of Conferences font
\usepackage{hyperref}
\usepackage{url}
\hypersetup{colorlinks=true,citecolor=blue,urlcolor=blue,linkcolor=blue}
\usepackage[utf8]{inputenc}

\usepackage{amssymb,amsmath,amsfonts}
\usepackage{mathtools}
\usepackage{mathrsfs}
\usepackage{bbm}
\usepackage{slashed}
\usepackage{nicefrac}
\usepackage{graphicx}
\usepackage{caption}
\usepackage{subcaption}
\usepackage{esint}
\usepackage{graphicx}
\usepackage[dvipsnames]{xcolor}
\usepackage{array}
\usepackage{simplewick}
\usepackage{hyperref}
\usepackage{xparse}
\usepackage{xspace}
\usepackage{tikz}
\usetikzlibrary{decorations.pathmorphing}
\usetikzlibrary{automata,positioning}
\usepackage{cancel}
\usepackage[normalem]{ulem}
\usepackage{xifthen}
\usepackage{dsfont}
\usepackage[titletoc]{appendix}
\usepackage{booktabs}
\usepackage{units}

\long\def\comment#1{ }

\def\be{\begin{eqnarray*}}
\def\ee{\end{eqnarray*}}
\def\beq{\begin{eqnarray}}
\def\eeq{\end{eqnarray}}

\def\0{{\boldsymbol 0}}

\def\x{{\boldsymbol x}}

\def\rmd{{\rm d}}

\def\and{ \quad\text{and}\quad}

\def\cK{{\cal K}}

\begin{document}
\title{Nonlinear dynamics of jet quenching}

\author{\firstname{Yacine} \lastname{Mehtar-Tani}\inst{1,3}\fnsep\thanks{\email{mehtartani@bnl.gov}}
}

\institute{Physics Department, Brookhaven National Laboratory, Upton, NY 11973, USA          }

\abstract{
We present an analytic framework for jet quenching in dense QCD matter that unifies medium--induced branching with vacuum collinear evolution. Energy transported outside the jet region is governed by a non--linear rate equation that resums arbitrary--angle gluon splittings, each enhanced by the medium length $L$.  We show that for asymptotically large energies the energy loss distribution for a single hard parton exhibits an exponential (generalized Poisson) behavior that  provides the initial condition for a non--linear, DGLAP--like evolution that resums collinear logarithms associated with early in-vacuum fragmentation and with the medium’s angular resolution. This framework allows a systematic resummations of parton branching contributions to jet energy loss. 
}
\maketitle

%%%%%%%%%%%%%%%%
\section{Introduction}
%%%%%%%%%%%%%%%%

The suppression of high-$p_T$ hadrons and fully reconstructed jets at RHIC and the LHC~(see Ref.~\cite{Blaizot:2015lma} for a review and references therein) has established jet quenching as one of the clearest signatures of the quark--gluon plasma. A theoretical description must consistently account for multiple scattering, medium--induced parton showers~\cite{Mehtar-Tani:2012mfa,Blaizot:2013vha,Caucal:2019uvr,Arnold:2020uzm,Mehtar-Tani:2021fud}, and the interplay of quantum interference and decoherence~\cite{Mehtar-Tani:2010ebp,Mehtar-Tani:2011hma,Casalderrey-Solana:2011ule,Mehtar-Tani:2017ypq}.  

Color coherence and its eventual loss play a central role in determining the fate of energetic jets. In vacuum, angular ordering ensures that successive emissions remain coherent until they are resolved at smaller and smaller angles. In a medium, multiple scatterings induce transverse momentum broadening that can disentangle the color correlation between branches. Once color decoherence sets in, daughter partons radiate independently, dramatically enhancing the efficiency of energy transport away from the jet core \cite{Mehtar-Tani:2017ypq}. The competition between coherence at small angles and decoherence at large angles defines the medium’s resolution scale, $\theta_c \sim (\hat q L^3)^{-1/2}$, which will be central in what follows.  

Radiative energy loss is the dominant mechanism behind jet quenching. Because of its non--local nature, gluon emission is sensitive to multiple scattering during the formation time $t_f\simeq \sqrt{\omega/\hat q}$. This interference leads to the well--known Landau--Pomeranchuk--Migdal (LPM) suppression of hard gluons and introduces the characteristic scale $\omega_c=\hat q L^2$~\cite{Baier:1996kr,Zakharov:1996fv}. At the same time, repeated emissions give rise to turbulent cascades that redistribute energy self--similarly toward the medium’s soft scales~\cite{Blaizot:2014ula}. Beyond inclusive energy loss, however, a complete understanding of quenching requires tracing how energy crosses the jet boundary. Radiation that leaves the jet cone may subsequently radiate back inside, leading to a subtle interplay that is reminiscent of non--global observables in vacuum QCD~\cite{Banfi:2002hw}. Substructure measurements and recent Monte Carlo studies have begun to expose these dynamics experimentally.  

The aim of this work \cite{Mehtar-Tani:2024mvl} is to construct a unified analytic description of out--of--cone energy flow that naturally incorporates the onset of decoherence, resums medium--induced branchings, and matches onto collinear vacuum evolution above the medium’s resolution angle. As a first step, we introduce the energy--loss distribution $S_{\rm loss}(\epsilon,E)$, which specifies the probability that a parton of initial energy $E$ loses an amount $\epsilon$ of energy to the medium. The inclusive jet spectrum can then be written as
\begin{equation}
\label{eq:factorization}
\frac{\rmd \sigma_{\rm jet}}{\rmd p_T}
= \int_0^\infty \rmd E \int_0^\infty \rmd \epsilon\,
\delta(p_T-E+\epsilon)\, S_{\rm loss}(\epsilon,E)\,
\frac{\rmd \sigma^{\rm vac}_{\rm jet}}{\rmd E}\,,
\end{equation}
which makes explicit how the medium correction enters through a convolution with the vacuum cross section. This factorized form will serve as the starting point for our analysis.  

%%%%%%%%%%%%%%%%%%%%%%%%%%%%%%
\section{Non--linear evolution of jet energy loss}
%%%%%%%%%%%%%%%%%%%%%%%%%%%%%%

The Laplace transform provides a natural representation of Eq.~\eqref{eq:factorization}, as it eliminates the convolution structure and renders the evolution local in $\nu$. At high energies, the medium contribution becomes nearly independent of $E$, which simplifies the analysis.  Hence, it is convenient to work with the quenching factor \cite{Baier:2001yt,Mehtar-Tani:2017web}
\beq
Q_\nu(E)=\int_0^\infty d\epsilon\, e^{-\nu\epsilon} S_{\rm loss}(\epsilon,E)\,.
\eeq

The physical picture is as follows. The medium resolves color charges only at angles larger than $\theta_c$, while at smaller angles radiation remains coherent. This divides the dynamics into two stages: a medium--induced cascade that resums $(\alpha_s L)^n$ contributions at arbitrary angles, and a vacuum--like collinear cascade that resums logarithms of $R$ and $R/\theta_c$. The matching between the two occurs at the medium resolution angle $\theta_c =(\hat q L^3)^{-1/2}$, marking the point where color decoherence sets in and partons begin to radiate independently \cite{Mehtar-Tani:2011hma,Casalderrey-Solana:2012evi,Mehtar-Tani:2017ypq,Mehtar-Tani:2017web}.  

Elastic broadening keeps the leading parton inside the cone with high probability when $E\gg \alpha_s^2 \omega_c$, so the dominant out--of--cone energy flow arises from radiation. Successive splittings create branches that, after decoherence, evolve independently and contribute additively to the energy flow. This independence is encoded in the non--linear master equation \cite{Mehtar-Tani:2024mvl},
\begin{equation}
\label{eq:master}
Q_\nu(t;E) \;=\; Q_\nu^{\rm el}(t;E)
+ \alpha_s \!\int_0^t\!\rmd t_1 \!\int_0^1\!\rmd z\;\cK(z,E)\,
\Big[ Q_\nu(t{-}t_1;zE)\,Q_\nu(t{-}t_1;(1{-}z)E)-Q_\nu(t{-}t_1;E)\Big],
\end{equation}
where $\cK(z,E)$ is the in-medium splitting rate~\cite{Blaizot:2013vha}. This equation resums arbitrary branchings with exact energy sharing, enforces unitarity through the subtraction term, and admits the thermal fixed point $Q_\nu^{\rm th}(E)=e^{-\nu E}$, which corresponds to total energy loss $S_{\rm loss}(\epsilon)=\delta(\epsilon-E)$.  The initial condition of the above evolution is given by $Q_\nu^{\rm el}(t;E)$ which encodes the physics of energy leakage out of the cone due to elastic scattering \cite{Mehtar-Tani:2024mvl}. 

In the  single-hard parton regime, where we neglect early collinear DGLAP evolution, Eq.~\eqref{eq:master} reduces to an exponential form in Laplace space, corresponding to a generalized Poisson distribution of energy loss. The independent--emission ansatz of Baier \textit{et al.}~\cite{Baier:2001yt} is thus recovered as a limiting case ($R\!\to\!0$). In the regime $R\gg \theta_c$, achieved in a dense or large media, the cascade matches to a non--linear DGLAP--like evolution that resums collinear logarithms, organizing large contributions in $\alpha_s \ln (R/\theta_c)$, and $(\alpha_s L)^n$ through the initial condition given by the solution to Eq.~\eqref{eq:master}. 

%%%%%%%%%%%%%%%%%%%%%%%%%%%%%%
\section{Summary and outlook}
%%%%%%%%%%%%%%%%%%%%%%%%%%%%%%
The framework makes explicit how $L$, $\hat q$, and $R$ govern the balance between in– and out–of–cone energy flow. Increasing either $L$ or $\hat q$ enhances the number of resolvable emissions and drives the system toward the thermal fixed point. In contrast, reducing $R$ increases the leakage of a single color charge but simultaneously decreases the number of effective emitters due to the vacuum collimator. The $R$ dependence therefore reflects the competition between these two effects. This interplay has been analyzed and confronted with data, which show only a weak $R$ dependence~\cite{Mehtar-Tani:2021fud}.

Because radiation may both leave and reenter the cone, the problem is inherently non--global. The nonlinear equation~\eqref{eq:master} captures this non--global structure in a probabilistic form: once decoherence occurs, daughter partons evolve independently and their energy losses add. This framework is directly applicable to substructure observables, including groomed jets and subjet multiplicities, and provides a natural platform for incorporating medium response at soft scales. In particular, the probabilistic picture offers a new handle on the role of soft radiation that may be absorbed by or returned from the medium, a long--standing open problem in connecting perturbative descriptions with hydrodynamic response.
Leading NLO corrections to $\hat q$ and to splitting kernels~\cite{Blaizot:2014bha} can be incorporated as renormalizations of the rate and broadening inputs without altering the structure of Eq.~\eqref{eq:master}. Time--dependent $\hat q(t)$ can account for expansion or inhomogeneities, while quark channels follow through simple color substitutions. Phenomenology can proceed by solving Eq.~\eqref{eq:master} with realistic medium profiles. Importantly, the master equation provides an analytic counterpart to Monte Carlo Event Generators such as JetMed \cite{Caucal:2019uvr}, clarifying their approximations and highlighting which higher--order effects are systematically included. This resummation program has recently been recast into a factorized effective field theory framework~\cite{Mehtar-Tani:2024smp}, enabling systematic higher-order calculations of jet observables.

In summary, we have developed a unified description of jet quenching built on three elements: a non--linear master equation for medium--induced cascades, a consistent matching to vacuum collinear evolution governed by decoherence, and a generalized Poisson structure that clarifies asymptotics and simplifies computation. This formulation provides a clear organizational principle for higher--order jet studies in heavy--ion collisions and a natural bridge to future phenomenology at the Electron--Ion Collider. Beyond its immediate applications, it opens the way to systematic precision studies of jet quenching, where controlled perturbative inputs are embedded into a framework flexible enough to incorporate medium dynamics. In this sense, the approach provides both a theoretical anchor for existing observations at the LHC and RHIC, and a roadmap for interpreting future measurements at high luminosity and new facilities, ultimately connecting the microscopic dynamics of parton cascades with macroscopic properties of hot and dense QCD matter.

%%%%%%%%%%%%%%%%%%%%%%
\section*{Acknowledgements}
%%%%%%%%%%%%%%%%%%%%%%

This work was supported by the U.S. Department of Energy under Contract No. DE-SC0012704 and by Laboratory Directed Research and Development (LDRD) funds from Brookhaven Science Associates.

\bibliography{QM25-mehtar-2}
%Journal Author, Article title. Journal \textbf{Volume}, page numbers (year). \url{https://doi.org/Article-DOI-number}
%Book Author, \textit{Book title} (Publisher, place, year) page numbers
%\end{thebibliography}

\end{document}